\documentclass[
twocolumn,
preprintnumbers,
nofootinbib,
prd,aps
]{revtex4}

\usepackage{epsf}
\usepackage{latexsym}
\usepackage{amssymb}
\usepackage{epsf}
\usepackage{amsmath}
\usepackage{slashed}

\usepackage{graphicx}
\usepackage{comment}

\begin{document}


\newcommand{\rem}[1]{{\bf #1}}

\renewcommand{\topfraction}{0.8}

\preprint{UT-17-26}

\title{

  State-of-the-Art Calculation of the Decay Rate of Electroweak Vacuum
  \\
  in Standard Model

}

\author{
So Chigusa$^{(a)}$, Takeo Moroi$^{(a)}$ and Yutaro Shoji$^{(b)}$
}

\affiliation{
$^{(a)}$Department of Physics, The University of Tokyo, Tokyo 113-0033, Japan
\\
$^{(b)}$Institute for Cosmic Ray Research, The University of Tokyo, 
Kashiwa 277-8582, Japan
}

\date{July, 2017}

\begin{abstract}

  The decay rate of the electroweak (EW) vacuum is calculated in the
  framework of the standard model (SM) of particle physics, using the
  recent progresses in the understanding of the decay rate of
  metastable vacuum in gauge theories.  We give a manifestly
  gauge-invariant expression of the decay rate.  We also perform a
  detailed numerical calculation of the decay rate.  With the best-fit
  values of the SM parameters, we find that the decay rate of the EW
  vacuum per unit volume is about $10^{-577}\ {\rm Gyr^{-1}Gpc^{-3}}$;
  with the uncertainty in the top mass, the decay rate is estimated as
  $10^{-295}-10^{-1465}\ {\rm Gyr^{-1}Gpc^{-3}}$.

\end{abstract}

\maketitle

\renewcommand{\thefootnote}{\#\arabic{footnote}}

\noindent\underline{\it Introduction:}
It is highly non-trivial whether the vacuum we are living in, which we
call electroweak (EW) vacuum, is absolutely stable or not.  If there
exists a vacuum which has lower energy density than that of the EW
vacuum, which is the case in a large class of particle-physics models,
the EW vacuum decays via the quantum tunneling effect.  If the decay
rate is too large, the universe should have been experienced a phase
transition before the present epoch, with which the universe would
show completely different aspects than the present one.  From the
particle-physics and cosmology points of view, the stability of the EW
vacuum is of particular interest to have deep insight into
particle-physics models and the nature of the universe.

Even in the standard model (SM) of particle physics, which is
extremely successful to explain particle interactions, the EW vacuum
may be metastable \cite{Isidori:2001bm, Degrassi:2012ry,
  Alekhin:2012py, Espinosa:2015qea, Plascencia:2015pga, Lalak:2016zlv,
  Espinosa:2016nld}.  In particular, the discovery of the Higgs boson
by the LHC experiments \cite{Aad:2012tfa, Chatrchyan:2012xdj} shed
light on the stability of the EW vacuum.  The observed value of the
Higgs mass suggests that the Higgs quartic coupling becomes negative
via the renormalization group (RG) effects at energy scale much higher
than the EW scale.  This fact implies that the Higgs potential becomes
negative and that the EW vacuum is not absolutely stable if the SM is
valid up to a scale much higher than the EW scale.

The decay rate of the EW vacuum has been estimated in the past, mostly
using the method given in \cite{Coleman:1977py, Callan:1977pt,
  Coleman:aspectsof}.  The decay rate of the metastable vacuum (i.e.,
false vacuum) per unit volume, which we call $\gamma$, is given in the
following form:
\begin{align}
  \gamma = {\cal A} e^{-{\cal B}},
\end{align}
where ${\cal B}$ is the action of the so-called bounce, which is the
solution of the four-dimensional (4D) Euclidean equation of motion,
while ${\cal A}$ takes account of the fluctuation around the bounce.
The bounce action ${\cal B}$ can be evaluated relatively easily, while
the calculation of the prefactor ${\cal A}$ is complicated both
conceptually and numerically.  In particular, if the bounce is coupled
to gauge fields, which is the case when considering the decay of the
EW vacuum in the SM, gauge-invariant calculation of ${\cal A}$ has not
have been performed.  In addition, in the calculation of the decay
rate of the EW vacuum, path integral of the zero-mode in association
with the (approximate) classical conformal invariance was not properly
performed.  The calculation of $\gamma$ in the past could not avoid
some or all of these difficulties, resulting in ambiguities in the
final result.

Recently, however, a new formalism has been developed to calculate
$\gamma$, which can give a manifestly gauge-invariant expression of
${\cal A}$ \cite{Endo:2017tsz, Endo:2017gal}.  By using the method
given there, a more unambiguous calculation of the decay rate of the
EW vacuum has become possible.

The main purpose of this letter is to perform a state-of-the-art
calculation of the decay rate of the EW vacuum in the framework of the
SM, using the recent progresses to calculate the decay rate of
metastable vacuum. We give a gauge-invariant expression of the decay
rate of the EW vacuum.  We also give a prescription to properly take
care of the zero-mode in association with the (classical) conformal
invariance, which shows up in the limit of large Higgs amplitude.
Then, we perform numerical calculation to estimate the decay rate, and
show that the decay rate for the size of the present Hubble volume is
much smaller than the inverse of the present age of the universe.  

\noindent\underline{\it Higgs potential and the bounce:}
In the following, we consider the situation where the Higgs potential
becomes negative due to the RG running of the quartic coupling of the
Higgs potential.  The instability of the potential occurs when the
Higgs amplitude becomes much larger than the EW scale; in the rest of
this letter, we concentrate on such a large Higgs amplitude.  Then,
denoting the Higgs doublet as $\Phi$, the Higgs potential is well
approximated by the quartic one \cite{Isidori:2001bm}:
\begin{align}
  V (\Phi) = \lambda (\Phi^\dagger \Phi)^2.
\end{align}
When the renormalization scale is relatively large, $\lambda<0$ is
realized.

For the study of the decay of the false vacuum, we first consider the
bounce, which corresponds to the classical path connecting the false
and true vacua.  In the present case, by using $SU(2)$ and $U(1)$
transformations, we can take the following bounce configuration:
\begin{align}
  \Phi |_{\rm bounce}  = \frac{1}{\sqrt{2}}
  \left( \begin{array}{c}
      0 \\ \bar{\phi} (r)
    \end{array} \right)
\end{align}
with vanishing gauge fields.  Here, $\bar{\phi}$ is a real function of
$r$ (with $r$ being the 4D radius in the Euclidean space) which obeys
\begin{align}
  \partial_r^2 \bar{\phi} + \frac{3}{r} \partial_r \bar{\phi}
  - \lambda \bar{\phi}^3 = 0,
\end{align}
with 
\begin{align}
  \partial_r \bar{\phi}(r=0) = 0,~~~
  \bar{\phi}(r=\infty) = 0.
\end{align}
Assuming that $\lambda<0$, the solution of the above equation is given
by
\begin{align}
  \bar{\phi} = 
  \bar{\phi}_C
  \left( 1 + \frac{|\lambda|}{8} \bar{\phi}_C^2 r^2 \right)^{-1},
\end{align}
where $\bar{\phi}_C$ is a constant which corresponds to the bounce
amplitude at the center of the bounce configuration.  Notice that the
bounce contains a free parameter $\bar{\phi}_C$.  The bounce action is
given by
\begin{align}
  {\cal B} = \frac{8\pi^2}{3|\lambda|}.
\end{align}

\noindent\underline{\it Decay rate:} Now we are at the position to
calculate the decay rate of the EW vacuum.  As we have shown, we
already have the analytic expression of the bounce action ${\cal B}$.
On the other hand, the calculation of the prefactor ${\cal A}$ is
highly non-trivial.  The prefactor ${\cal A}$ is obtained by
calculating the functional determinants of the fluctuation operators
of the fields that couple to the bounce field \cite{Callan:1977pt}.

First, let us consider the (physical) Higgs field $h$, which is
embedded into the Higgs doublet as
\begin{align}
  \Phi  = \frac{1}{\sqrt{2}}
  \left( \begin{array}{c}
      \varphi^{1} + i\varphi^{2}
      \\
      \bar{\phi} (r) + h + i\varphi^{3}
    \end{array} \right),
\end{align}
where $\varphi^a$ are Nambu-Goldstone (NG) modes.  After the
decomposition with respect to the 4D angular-momentum, the fluctuation
operator of $h$ is given by
\begin{align}
  {\cal M}_J^{(h)} = 
  -\Delta_J
  - 3 |\lambda| \bar{\phi}^2,
\end{align}
where $J$ characterizes the eigenvalues of 4D angular-momentum
operators, and it takes $J=0$, $\frac{1}{2}$, $1$, $\cdots$, with
which $\mbox{Det}{\cal M}^{(h)}=\prod_J [\mbox{Det}{\cal
  M}_J^{(h)}]^{(2J+1)^2}$.  In addition,
\begin{align}
  \Delta_J\equiv
  \partial_r^2 + \frac{3}{r} \partial_r - \frac{4J(J+1)}{r^2}.
\end{align}

The ratio of the functional determinants of the fluctuation operators
relevant for the calculation of ${\cal A}$ can be performed with the
method given in \cite{Coleman:aspectsof, Dashen:1974ci,
  Kirsten:2003py, Kirsten:2004qv}.  For the calculation, we first
limit the region as $0\leq r\leq r_\infty$, where $r_\infty$ is a
(large) radius which is taken to be infinity at the end of
calculation, and impose relevant boundary conditions for the mode
functions at $r=0$ and $r=r_\infty$.  By using analytic properties of
the functional determinants, we obtain
\begin{align}
  \frac{\mbox{Det}{\cal M}_J^{(h)}}{\mbox{Det}\widehat{\cal M}_J^{(h)}}
  = 
  \frac{f_J^{(h)}(r_\infty)}{r_\infty^{2J}},
  \label{detM/DetMhat}
\end{align}
where $\widehat{\cal M}_J^{(h)}=[{\cal
  M}_J^{(h)}]_{\bar{\phi}\rightarrow 0}$ is the fluctuation operator
around the false vacuum, and $f_J^{(h)}$ obeys
\begin{align}
  {\cal M}_J^{(h)} f_J^{(h)} = 0,
\end{align}
with the boundary condition $f_J^{(h)}(r\rightarrow 0)\simeq r^{2J}$.

For $J\geq 1$, the functional determinants necessary for the
calculation of ${\cal A}$ are obtained by using Eq.\
\eqref{detM/DetMhat}.  On the contrary, for $J=0$ and $J=\frac{1}{2}$,
special care is needed because of the existence of zero-modes; ${\cal
  A}$ diverges if one naively uses Eq.\ \eqref{detM/DetMhat} for those
cases.

The zero-mode for $J=0$ is related to the conformal invariance; in the
present analysis, we approximate that the Higgs potential is quartic,
and hence the theory has a conformal invariance at the classical
level.  Consequently, the bounce configuration is not uniquely
determined and its continuous deformation with respect to the
parameter $\bar{\phi}_C$ is possible.  This is easily understood from
the expression of the mode function of the conformal zero-mode, which
is given by
\begin{align}
  \psi^{\rm (conf)}\equiv  \,&
  {\cal N}_{\rm conf}
  \left( 1 - \frac{|\lambda|}{8} \bar{\phi}_C^2 r^2 \right)
  \left( 1 + \frac{|\lambda|}{8} \bar{\phi}_C^2 r^2 \right)^{-2}
  \nonumber \\ = \,&
  {\cal N}_{\rm conf}
  \frac{\partial\bar{\phi}}{\partial\bar{\phi}_C},
\end{align}
where ${\cal N}_{\rm conf}$ is the normalization factor.  Indeed,
one can see that ${\cal M}^{(h)}_0 \psi^{\rm (conf)}=0$.  The
normalization factor is given by
\begin{align}
  {\cal N}_{\rm conf}^{-2}
  = 
  \frac{1}{2\pi} \int d^4 r 
  \left( \frac{\partial\bar{\phi}}{\partial\bar{\phi}_C} \right)^2
  \simeq 
  \frac{64\pi}{|\lambda|^2 \bar{\phi}_C^4}
  \ln r_\infty.
\end{align}
We comment here that ${\cal N}_{\rm conf}^{-2}$ diverges when
$r_\infty$ is taken to infinity.  As we will see below, however,
${\cal N}_{\rm conf}$ disappears from the final expression by properly
taking into account the measure of the path integral of the conformal
zero-mode.

Because the zero-mode wave function in association with the conformal
invariance is given by the derivative of $\bar{\phi}$ with respect to
$\bar{\phi}_C$, the path integral of the conformal zero-mode should be
regarded as the integration over all the possible deformation of the
bounce configuration with the change of $\bar{\phi}_C$:
\begin{align}
  \int {\cal D}h^{\rm (conf)}
  \rightarrow 
  \int \frac{d\bar{\phi}_C}{{\cal N}_{\rm conf}}.
\end{align}
Then, remembering that the functional determinants originate from the
path integral of the fields coupled to the bounce, the functional
determinant of ${\cal M}^{(h)}_0$ should be understood as
\begin{align}
  \left[ \mbox{Det} {\cal M}^{(h)}_0 \right]^{-1/2}
  \rightarrow
  \int \frac{d\bar{\phi}_C}{{\cal N}_{\rm conf}}
  \left[ \mbox{Det}' {\cal M}^{(h)}_0 \right]^{-1/2},
\end{align}
where the ``prime'' indicates that the zero eigenvalue is omitted from
the functional determinant.  In order to omit the zero eigenvalue, we
use the following technique \cite{Endo:2017tsz}:
\begin{align}
  \frac{\mbox{Det}'{\cal M}^{(h)}_0}
  {\mbox{Det}\widehat{\cal M}^{(h)}_0}
  = 
  \lim_{\nu\rightarrow 0} \nu^{-1}
  \frac{\mbox{Det}({\cal M}^{(h)}_0+\nu)}
  {\mbox{Det}\widehat{\cal M}^{(h)}_0}
  = 
  \check{f}^{(h)}_0 (r_\infty),
\end{align}
where the function $\check{f}^{(h)}_0$ satisfies
\begin{align}
  \left( 
    \partial_r^2 + \frac{3}{r} \partial_r 
    + 3 |\lambda| \bar{\phi}^2
  \right) \check{f}^{(h)}_0 =
  \frac{\partial\bar{\phi}}{\partial\bar{\phi}_C},
\end{align}
and $\check{f}^{(h)}_0(r\rightarrow 0)=0$, resulting in
\begin{align}
  \check{f}^{(h)}_0 (r_\infty) 
  = \int_0^{r_\infty} dr_1 r_1^{-3}
  \int_0^{r_1} dr_2 r_2^3 
  \frac{\partial\bar{\phi}}{\partial\bar{\phi}_C}
  \simeq
  -\frac{4}{|\lambda| \bar{\phi}_C^2} \ln r_\infty.
\end{align}
Consequently,
\begin{align}
  \left|
    \frac{\mbox{Det} {\cal M}^{(h)}_0}{\mbox{Det} \widehat{\cal M}^{(h)}_0}
  \right|^{-1/2}
  \rightarrow
  \int \frac{d\bar{\phi}_C}{\bar{\phi}_C}
  \left( \frac{16\pi}{|\lambda|} \right)^{1/2}.
\end{align}

The zero-modes for $J=\frac{1}{2}$ are related to the translational
invariance; they can be taken care of as \cite{Callan:1977pt}
\begin{align}
  \frac{\mbox{Det}{\cal M}^{(h)}_{1/2}}
  {\mbox{Det}\widehat{\cal M}^{(h)}_{1/2}}
  \rightarrow
  {\cal V}_{\rm 4D}^{-1/2}
  \left( \frac{{\cal B}}{2\pi} \right)^{-1}
  \frac{\check{f}^{(h)}_{1/2} (r_\infty)}{r_\infty},
\end{align}
where ${\cal V}_{\rm 4D}$ is the volume of the 4D Euclidean space, and
the function $\check{f}^{(h)}_{1/2}$ obeys
\begin{align}
  {\cal M}_{1/2}^{(h)} \check{f}^{(h)}_{1/2} = 
  - r \left( 1 + \frac{|\lambda|}{8} \bar{\phi}_C^2 r^2 \right)^{-2},
\end{align}
with $\check{f}^{(h)}_{1/2}(r\rightarrow 0)=0$.  Notice that
$\check{f}^{(h)}_{1/2}(r_\infty)\propto\bar{\phi}_C^{-2}$.

For the effects of the gauge- and NG-bosons, a new technique has been
recently developed in \cite{Endo:2017gal, Endo:2017tsz}, which gives a
simple and manifestly gauge-invariant formula for the gauge- and
NG-boson contributions.  In \cite{Endo:2017gal, Endo:2017tsz}, the
scalar potential was assumed to be quadratic around the false vacuum,
while it is quartic in the present case.  Based on
\cite{Endo:2017tsz}, we derive the formula relevant for the present
case.  (The result is given in Eq.\ \eqref{I(WZNG)}; a more detailed
derivation of the following formulae will be given elsewhere
\cite{ChiMorSho}.)

Combining the contributions of particles which have sizable couplings
with the bounce, the decay rate of the EW vacuum is expressed as
\begin{align}
  \gamma = 
  \int d\ln\bar{\phi}_C
  \left[
    I^{(h)}
    I^{(W,Z,{\rm NG})}
    I^{(t)}
    e^{-\delta{\cal S}_{\overline{\rm MS}}}
    e^{-{\cal B}}
  \right]_{\mu(\bar{\phi}_C)},
  \label{gamma}
\end{align}
where $\delta{\cal S}_{\overline{\rm MS}}$ is the effect of the
so-called divergent part \cite{Endo:2017tsz} (which is calculated with
the $\overline{\rm MS}$ scheme), and $\mu$ is the renormalization
scale at which the SM coupling constants for the calculation of the
integrand are evaluated.  The Higgs contribution as well as the gauge-
and NG-boson contribution are given by
\begin{widetext}
\begin{align}
  I^{(h)} = \, &
  \frac{{\cal B}^2}{4\pi^2}
  \left( \frac{16\pi}{|\lambda|} \right)^{1/2}
  \left[
    \frac{\check{f}^{(h)}_{1/2}(r_\infty)}{r_\infty}
  \right]^{-2}
  e^{s_0^{(h)}+s_{1/2}^{(h)}}
  \prod_{J\geq 1}
  e^{s_J^{(h)}}
  \left[
    \frac{f^{(h)}_J(r_\infty)}{r_\infty^{2J}}
  \right]^{-(2J+1)^2/2},
  \\
  I^{(W,Z,{\rm NG})} =  \, &
  {\cal V}_{SU(2)} 
  \left( \frac{16\pi}{|\lambda|} \right)^{3/2}
  \prod_{V=W^1,W^2,Z}
  e^{s_0^{(V,{\rm NG})}}
  \prod_{J\geq 1/2}
  e^{s_J^{(V,{\rm NG})}}
  \left[ 
    \frac{|\lambda| J \bar{\phi}_C^2 f_J^{(\eta^V)}(r_\infty)}
    {8 (J+1) r_\infty^{2J-2}}
  \right]^{-(2J+1)^2/2}
  \left[
    \frac{f^{(T^V)}_J(r_\infty)}{r_\infty^{2J}}
  \right]^{-(2J+1)^2},
  \label{I(WZNG)}
\end{align}
\end{widetext}
where ${\cal V}_{SU(2)}=2\pi^2$ is the volume of the $SU(2)$ group
parameterizing the possible deformation of the bounce configuration.
Here, $s_J^{(h)}$ and $s_J^{(V,{\rm NG})}$ are the effects of
counterterms to subtract divergences; the calculations of these
quantities are found in \cite{Isidori:2001bm, Endo:2017tsz}.  The
functions $f^{(\eta^V)}_J$ and $f^{(T^V)}_J$ satisfy
\begin{align}
  &
  (\Delta_J - g_V^2 \bar{\phi}^2) f^{(\eta^V)}_J
  - \frac{2 \bar{\phi}'}{r^2 \bar{\phi}} \partial_r
  \left(
    r^2 f^{(\eta^V)}_J
  \right)
  = 0,
  \label{eq_eta}
  \\ &
  (\Delta_J - g_V^2 \bar{\phi}^2) f^{(T^V)}_J = 0,
  \label{eq_T}
\end{align}
and $f^{(\eta^V)}_J(r\rightarrow 0)\simeq f^{(T^V)}_J(r\rightarrow
0)\simeq r^{2J}$, where
\begin{align}
  g_V =
  \frac{1}{2} \times 
  \left\{
    \begin{array}{ll}
      g_2 & : V=W^1, W^2 \\[2mm]
      \sqrt{g_2^2 + g_1^2} & : V=Z
    \end{array} \right. ,
\end{align}
with $g_2$ and $g_1$ being the gauge coupling constants of $SU(2)_L$
and $U(1)_Y$, respectively.  Expression of the top contribution
$I^{(t)}$ can be found in \cite{Isidori:2001bm}.  We emphasize that
the above expressions for the decay rate is manifestly gauge
invariant; they hold irrespective of the choice of the gauge parameter
(which is often called $\xi$).  Furthermore, $\delta{\cal
  S}_{\overline{\rm MS}}$ is given by the sum of the Higgs and top
contributions as well as the gauge and NG contributions: $\delta{\cal
  S}_{\overline{\rm MS}}= \delta{\cal S}_{\overline{\rm MS}}^{(h)}+
\delta{\cal S}_{\overline{\rm MS}}^{(t)}+
\sum_{V=W^1,W^2,Z}\delta{\cal S}_{\overline{\rm MS}}^{(V,{\rm NG})}$.
The Higgs and top contributions are given in \cite{Isidori:2001bm},
while $\delta{\cal S}_{\overline{\rm MS}}^{(V,{\rm NG})}$ is obtained
with the prescription given in \cite{Endo:2017tsz}:
\begin{widetext}
\begin{align}
  \delta{\cal S}_{\overline{\rm MS}}^{(V,{\rm NG})} = 
  -\left(
    \frac{1}{3} + \frac{2 g_V^2}{|\lambda|} + \frac{g_V^4}{|\lambda|^2}
  \right)
  \left[
    \frac{5}{6} + \gamma_{\rm E} + 
    \ln \left( \sqrt{\frac{2}{|\lambda|}} \frac{\mu}{\bar{\phi}_C} \right)
  \right] 
  - \frac{2 g_V^2}{3 |\lambda|} + \frac{g_V^4}{3|\lambda|^2},
\end{align}
\end{widetext}
with $\gamma_{\rm E}$ being the Euler's constant.

In Eq.\ \eqref{gamma}, the renormalization scale $\mu$ is taken to be
$\bar{\phi}_C$-dependent in the following reason.  For fixed
$\bar{\phi}_C$, the typical mass scale of the fields which have
sizable couplings to the bounce is $O(\bar{\phi}_C)$, and only the
scales in the calculation are $\bar{\phi}_C$ and $\mu$.  Thus, one-loop
effects give terms proportional to $\ln (\bar{\phi}_C/\mu)$ to the
integrand; the $\mu$-dependence from such terms should be canceled by
the $\mu$-dependence of the coupling constants \cite{Endo:2015ixx}.  The
two- and higher-loop effects are expected to introduce terms
proportional to $\ln^p (\bar{\phi}_C/\mu)$ (with $p\geq 1$) which are,
on the contrary, not included in the present result.  In order to
minimize the higher order effects, we set
$\mu(\bar{\phi}_C)\sim\bar{\phi}_C$; hereafter, we take
$\mu(\bar{\phi}_C)=\bar{\phi}_C$ unless otherwise stated.  In fact, a
proper choice of $\mu$ is important for the convergence of the integral
over $\ln\bar{\phi}_C$.  In the SM, $\lambda$ is minimized at $\mu\sim
O(10^{17})\ {\rm GeV}$, and it increases above such a scale.  (The
runnings of the SM coupling constants are precisely included in our
numerical calculation; see the discussion below.)  Then, with taking
$\mu=\bar{\phi}_C$, because ${\cal B}$ is inversely proportional to
$|\lambda|$, the integrand of Eq.\ \eqref{gamma} is maximized when
$\bar{\phi}_C\sim O(10^{18})\ {\rm GeV}$ and is significantly
suppressed when $\bar{\phi}_C\gg O(10^{18})\ {\rm GeV}$.  Based on
this observation, we expect that the integration over $\bar{\phi}_C$
converges.

\noindent\underline{\it Numerical results:}
Now we apply our formula for the estimation of the decay rate of the
EW vacuum.  We evaluate $f^{(h)}_J$, $\check{f}^{(h)}_{1/2}$,
$f^{(\eta^V)}_J$, and $f^{(T^V)}_J$ (as well as other functions
necessary to calculate $I^{(t)}$ and counter terms) by numerically
solving differential equations.  The renormalization-scale dependence
of the SM coupling constants are evaluated by using the method given
in \cite{Buttazzo:2013uya}, which partially takes into account three-
and four-loop effects.  Then, with performing the integration over
$\ln\bar{\phi}_C$ numerically, the decay rate of the EW vacuum is
obtained.  We use the following Higgs and top masses
\cite{Olive:2016xmw}:
\begin{align}
  m_h =  \,&
  125.09 \pm 0.24 \ {\rm GeV},
  \label{higgsmass}
  \\
  m_t^{\rm (pole)} = \,&
  173.1 \pm 1.1 \ {\rm GeV},
  \label{topmass}
\end{align}
while the strong coupling constant is
\begin{align}
  \alpha_s (m_Z) = 0.1181\pm 0.0011.
  \label{alpha_S}
\end{align}

\begin{figure}[t]
  \centerline{\epsfxsize=0.475\textwidth\epsfbox{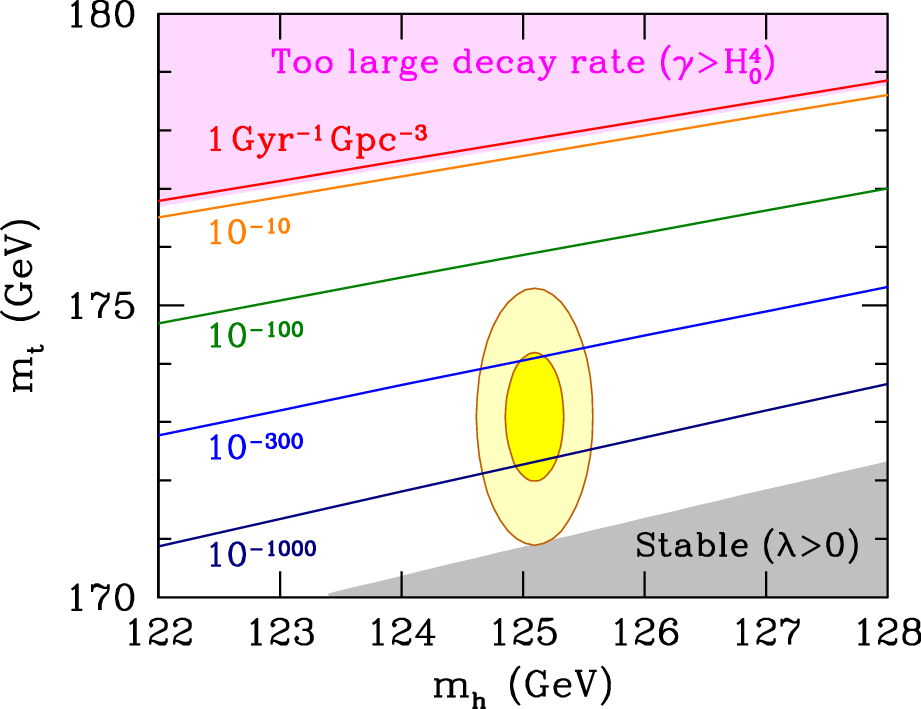}}
  \caption{\small The contours of constant $\gamma$ on Higgs mass vs.\
    top mass plane with $\alpha_s (m_Z) = 0.1181$.  The contours are
    $\gamma=1$, $10^{-10}$, $10^{-100}$, $10^{-300}$, and $10^{-1000}
    \ {\rm Gyr^{-1}Gpc^{-3}}$ from above.  In the upper shaded region
    (pink), $\gamma$ becomes larger than $H_0^4$.  In the lower shaded
    region (gray), the EW vacuum is stable because $\lambda$ is always
    positive.  We also show the constraint on the Higgs and top masses
    (yellow-shaded regions) adding their $1\sigma$ (inside) or
    $2\sigma$ (outside) uncertainties in quadrature.}
  \label{fig:gmm}
\end{figure}

For the best-fit values of the Higgs mass, top mass, and strong
coupling constant given above, we find $\gamma \simeq 10^{-741}\ {\rm
  GeV}^4 \simeq 10^{-577}\ {\rm Gyr^{-1}Gpc^{-3}}$.  Taking account of
the uncertainties, we obtain
\begin{align}
  \log_{10} [ \gamma\ ({\rm Gyr^{-1}Gpc^{-3}}) ]
  \simeq
  -577 {}_{-44}^{+40} {}_{-887}^{+283} {}_{-215}^{+143},
  \label{gamma577}
\end{align}
where the first, second, and third errors are due to those in the
Higgs mass, top mass, and the strong coupling constant given in Eqs.\
\eqref{higgsmass}, \eqref{topmass}, and \eqref{alpha_S}, respectively.
Thus, the decay rate is extremely sensitive to the top mass.  So far,
we have chosen the renormalization scale to be
$\mu(\bar{\phi}_C)=\bar{\phi}_C$.  Varying the renormalization scale
from $\mu(\bar{\phi}_C)=\frac{1}{2}\bar{\phi}_C$ to $2\bar{\phi}_C$,
for example, the change of the decay rate is
$\delta\log_{10}\gamma\sim 6$.  In Fig.\ \ref{fig:gmm}, we show the
contours of constant $\gamma$ on Higgs mass vs.\ top mass plane.

Comparing the decay rate with $H_0^{-4}\sim 10^{3}\ {\rm Gyr}{\rm
  Gpc}^3$ (with $H_0$ being the Hubble constant), the probability of
having a phase transition within the present Hubble volume for the
present cosmic time scale is enormously small for the best-fit values
of the SM parameters.  (Even if we vary $m_h$, $m_t^{\rm (pole)}$, and
$\alpha_s (m_Z)$ within $2\sigma$ uncertainties, $\gamma$ is at most
$10^{-68}\ {\rm Gyr}^{-1}{\rm Gpc}^{-3}$ which is still much smaller
than $H_0^4$.)  If the top mass were much larger than the observed
value, $\gamma$ would be larger than $\sim H_0^{4}$ so that the EW
vacuum would decay before the present epoch; such an instability bound
derived from our formula is consistent with that given in previous
work \cite{Degrassi:2012ry}.  In future, the universe will be
dominated by the dark energy, assuming that it is cosmological
constant.  Using the observed energy density of the dark energy, the
expansion rate will eventually become $H_{\infty}\simeq 56.3\ {\rm
  km/sec/Mpc}$ \cite{Ade:2015xua}.  Then, the phase transition rate
within the horizon scale of such a de Sitter universe is about
$10^{-575}\ {\rm Gyr}^{-1}\simeq 10^{-574} H_{\infty}$, which we
regard as the decay rate of the EW vacuum.  Uncertainty in this
estimation can be obtained from Eq.\ \eqref{gamma577}.

\noindent\underline{\it Summary:}
We have calculated the decay rate of the EW vacuum, assuming that the
SM is valid up to high energy scale.  We have derived a
gauge-invariant expression of the decay rate, properly performing the
path integral of the zero-mode in association with the conformal
invariance.  With the best-fit values of the Higgs and top masses and
$\alpha_s (m_Z)$, the decay rate of the EW vacuum per unit volume is
given by $10^{-577}\ {\rm Gyr}^{-1}{\rm Gpc}^{-3}$.  The probability
of the phase transition within the present horizon scale is found to
be enormously small.  This is a good news for us all because we can
safely live in the EW vacuum unless a new physics beyond the SM
significantly alters this conclusion.

\noindent{\it Acknowledgments:} This work was supported by the
Grant-in-Aid for Scientific Research C (No.26400239), and Innovative
Areas (No.16H06490).  The work of S.C. was also supported in part by
the Program for Leading Graduate Schools, MEXT, Japan.

\noindent{\it Note Added:} While preparing the manuscript, the paper
\cite{arXiv:1707.08124} showed up, which has significant overlap with
our work.  We found, however, several disagreements between the
results in \cite{arXiv:1707.08124} and ours, which are in (i) the
counterterms based on the angular-momentum decomposition
(corresponding to $s_J^{(V,{\rm NG})}$ in our calculation), (ii)
$\delta{\cal S}_{\overline{\rm MS}}^{(V,{\rm NG})}$, and (iii) the
volume of $SU(2)$ group.  Because of these, $\log_{10}\gamma$ based on
\cite{arXiv:1707.08124} becomes larger than ours by $\sim 65$.  In
addition, the method of the path integral over the conformal mode and
the choice of the renormalization scale are different; they result in
the shift of $\log_{10}\gamma$ by $\sim -33$, which should be regarded
as a theoretical uncertainty.

\newpage

~

\newpage

\renewcommand{\theequation}{A.\arabic{equation}}
\renewcommand{\thefigure}{A.\arabic{figure}}

\section*{\large Addendum: 2023 Update}
\setcounter{equation}{0}
\setcounter{figure}{0}

After the publication of this article, there have been improvements in
the determinations of SM parameters, particularly the Higgs boson
mass, the top-quark mass, and the strong coupling constants.  Thus, we
update our analysis using the latest values of those parameters.

\begin{figure}[t]
  \centerline{\epsfxsize=0.475\textwidth\epsfbox{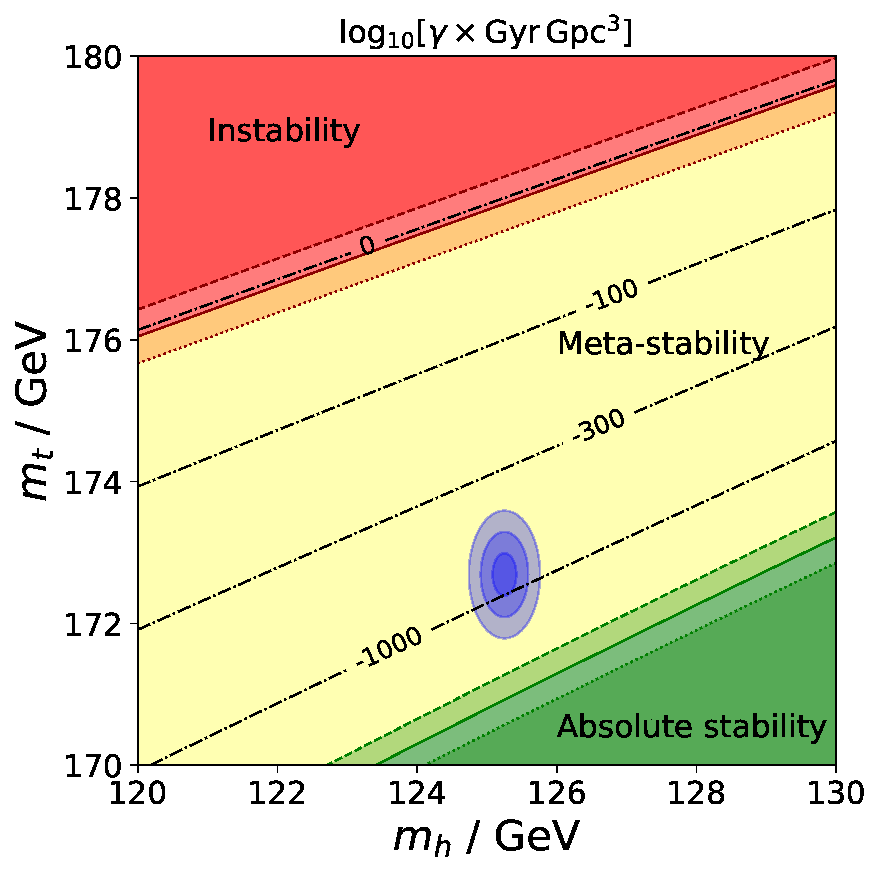}}
  \caption{\small The contours of constant $\gamma$ on Higgs mass
    vs.\ top mass plane, based on 2023 PDG values of SM
    parameters. The black dot-dashed contours are for $\gamma=1$,
    $10^{-100}$, $10^{-300}$, and $10^{-1000} \ {\rm
      Gyr^{-1}Gpc^{-3}}$, taking $\alpha_s (m_Z)$.  In the upper
    shaded region (pink), $\gamma$ becomes larger than $H_0^4$.  In
    the lower shaded region (green), the EW vacuum is stable because
    $\lambda$ is always positive.  Dependences of the boundaries of
    those regions are shown with lighter and darker colors.  We also
    show the constraint on the Higgs and top masses (blue-shaded
    regions) adding their $1\sigma$ (inside) $2\sigma$ (middle), and
    $3\sigma$ (outside) uncertainties in quadrature.}
  \label{fig:gmm2023}
\vspace{7mm}
  \centerline{\epsfxsize=0.475\textwidth\epsfbox{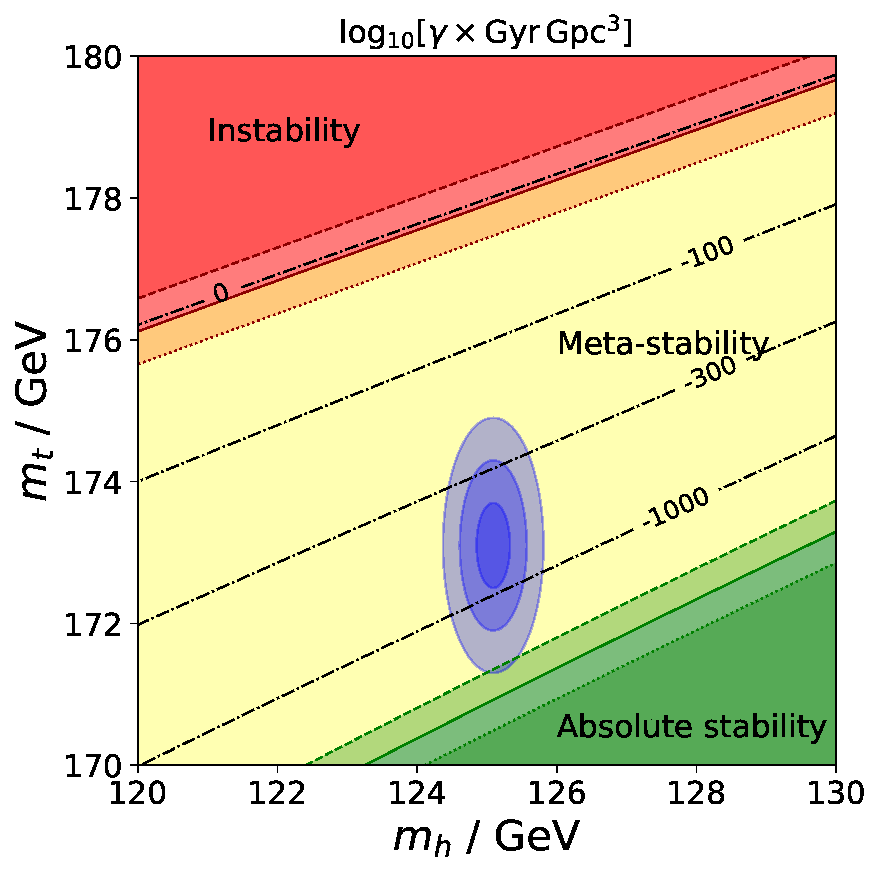}}
  \caption{\small The contours of constant $\gamma$ on Higgs mass
    vs.\ top mass plane, based on 2017 PDG values.}
  \label{fig:gmm2017}
\end{figure}

Based on the 2023 Particle Data Group (PDG) analysis
\cite{ParticleDataGroup:2022pth}, we adopt
\begin{align}
  m_h = \,& 125.25 \pm 0.17\, {\rm GeV},
  \\
  m_t = \,& 172.69 \pm 0.30\, {\rm GeV},
  \\
  \alpha_s (m_Z) =  \,& 0.1179 \pm 0.0009.
\end{align}
We calculate the decay rate of the EW vacuum in the SM using the {\tt
  ELVAS} package \cite{ELVAS}.  The contours of constant $\gamma$ is
shown in Fig.\ \ref{fig:gmm2023}.  Here, following
Ref.\ \cite{Chigusa:2018uuj}, we take the renormalization scale $\mu$
to be equal to the inverse of the bounce size $R^{-1}$ instead of
$\bar{\phi}_C$.  The best-fit value of $\gamma$ is $10^{-785}\, {\rm
  Gyr^{-1}Gpc^{-3}}$, while, with varying the SM parameters within the
$1\sigma$ ranges, we find
\begin{align}
  \log_{10} [ \gamma\ ({\rm Gyr^{-1}Gpc^{-3}}) ]
  \simeq
  -785 {}_{-50}^{+45} {}_{-222}^{+155} {}_{-277}^{+182},
\end{align}
where the first, second, and third errors are due to those in the
Higgs mass, top quark mass, and the strong coupling constant.

For comparison, in Fig.\ \ref{fig:gmm2017}, we also show the result
with the 2017 PDG values of the SM parameters
(particularly, for the top quark mass, 
here, we take the top quark mass from the direct measurements)
\cite{ParticleDataGroup:2016lqr}:
\begin{align}
  m_h^{(2017)} = \,& 125.09 \pm 0.24\, {\rm GeV},
  \\
  m_t^{(2017)} = \,& 173.1 \pm 0.6\, {\rm GeV},
  \\
  \alpha_s (m_Z)^{(2017)} =  \,& 0.1181 \pm 0.0011,
\end{align}
and $\mu=R^{-1}$.

\end{document}